\documentstyle[sprocl,epsf]{article}

\def\be{\begin{equation}}
\def\ee{\end{equation}}
\def\bea{\begin{eqnarray}}
\def\eea{\end{eqnarray}}

\begin{document}

\title{RESULTS FROM THE PHOBOS EXPERIMENT ON Au+Au COLLISIONS AT RHIC}
\author{KRZYSZTOF WO\,ZNIAK}
\address{Institute of Nuclear Physics  \\
ul. Kawiory 26A, 30-055 Krak\'ow, Poland \\
E-mail: krzysztof.wozniak@ifj.edu.pL} 
\author{for the PHOBOS Collaboration \\
B.B.Back$^1$, M.D.Baker$^2$,
D.S.Barton$^2$, S.Basilev$^5$, R.Baum$^8$,R.R.Betts$^{1,7}$, A.Bia\l as$^4$, 
R.Bindel$^8$, W.Bogucki$^3$, 
A.Budzanowski$^3$
% ...
}
\address{$^1$~Physics Division, Argonne National Laboratory,
$^2$~Chemistry and C-A Departments, Brookhaven National Laboratory, 
$^3$~Institute of Nuclear Physics, Poland,
$^4$~Department of Physics, Jagellonian University, Poland,
$^5$~Laboratory for Nuclear Science, Massachusetts Institute of Technology, 
$^6$~Department of Physics, National Central University, Taiwan, 
$^7$~Department of Physics, University of Illinois at Chicago, 
$^8$~Department of Chemistry, University of Maryland, 
$^9$~Department of Physics and Astronomy, University of Rochester, 
$^{10}$~Department of Physics, Yale University}


\maketitle\abstracts{ 
PHOBOS is one of four experiments studying the Au-Au interactions
at RHIC. The data collected during the first few weeks after
the RHIC start-up, using the initial configuration
of the PHOBOS detector, were sufficient to obtain the first
physics results for the most central collisions of Au nuclei
at the center of mass energy of 56 and 130 AGeV. The pseudorapidity
density of charged particles near midrapidity is
shown and compared with data at lower energies and from
$pp$ and $p\overline{p}$ collisions. The progress of the 
analysis of the data is also presented.
}

\section{Introduction}
In June 2000 the Relativistic Heavy Ion Collider (RHIC) 
in Brookhaven National Laboratory became operational 
and delivered first collisions of gold nuclei at the center of mass energy
several times larger than that available from accelerators so
far. This opened a new frontier in particle physics, with
a potential to study the behavior of strongly interacting
matter under the conditions of extreme energy densities.
Theoretical predictions of creation of the Quark-Gluon Plasma (QGP)
define the main goal for all RHIC experiments: search for this 
new state of matter.
  
PHOBOS experiment was designed to measure global properties of 
the Au+Au interactions (multiplicity and angular distribution 
of charged particles in 
almost full solid angle) and combine them with detailed data on 
particles emitted at large angles, which may carry information on the 
early stage of the interaction. This enables to correlate the total
multiplicity and centrality of the collision with several 
proposed QGP signals (particles ratios, transverse momentum distribution,
fluctuations).   

\section{The PHOBOS experiment}
The PHOBOS detector\ \cite{phobos1} consists of three main subsystems: 
the multiplicity detector,
the spectrometer extended by a time of flight wall (TOF)
and the trigger system (Fig.\ 1).

The multiplicity detector, covering almost full solid angle,
 is divided into octagonal, barrel like, 
part surrounding the nominal interaction point and six rings placed
along the beam pipe on both sides of the interaction point. 
Above and below the nominal interaction point a part of the octagonal
multiplicity detector is substituted by top and bottom vertex detectors
used to precise measure the interaction point.

The two arm spectrometer consists of several layers of silicon sensors 
placed inside a conventional magnet (2 Tesla field).
It covers pseudorapidity range approximately between 0 and 2 and 
about 10$^{\circ}$ in azimuth (each arm). Using the information 
from the spectrometer on particles trajectories and the ionization energy 
losses we can measure momenta 
of charged particles and identify them.

The trigger system consists of two identical sets of counters placed on 
both sides of the interaction point. 
The scintillator paddle counters (PP and PN) detect charged particles emitted 
in pseudorapidity range 3-4.5 and 
the Zero Degree Calorimeters (ZDCP and ZDCN) measure the energy deposited 
by spectator neutrons from the colliding nuclei.

\begin{figure}[t]
\begin{minipage}{5.8cm}
\epsfxsize=5.79cm\epsfbox{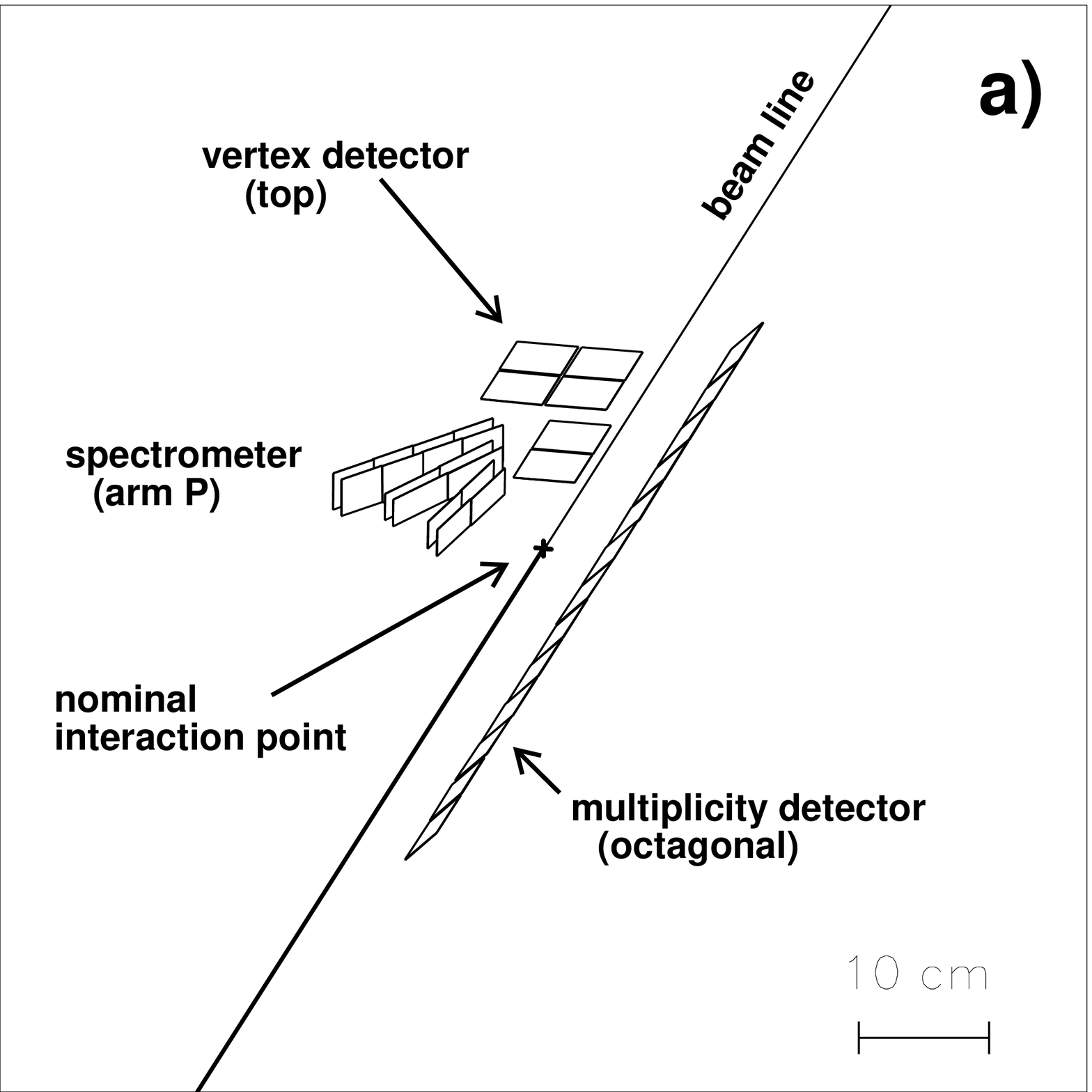}
\end{minipage} {\ }
\begin{minipage}{5.8cm}
\epsfxsize=5.79cm\epsfbox{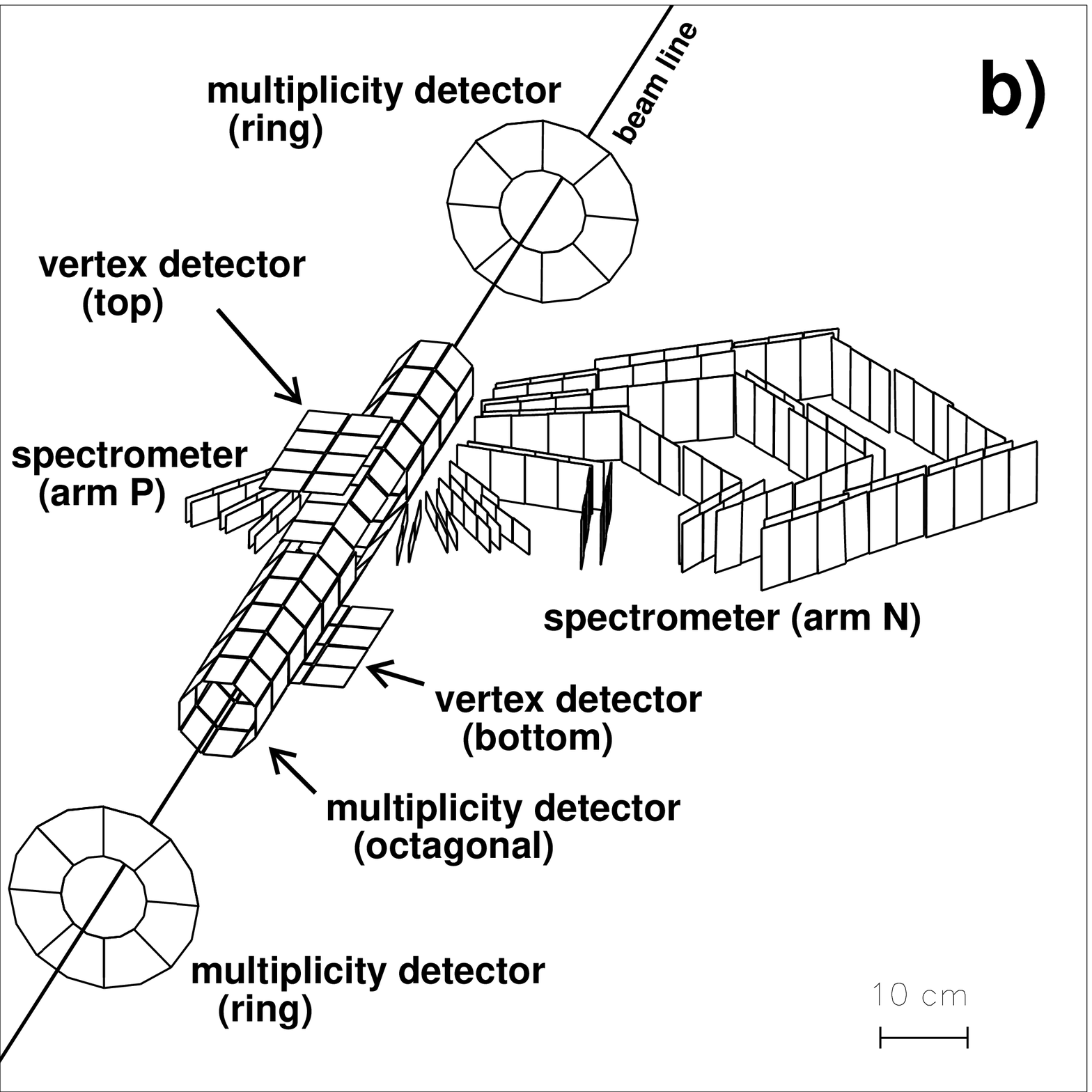}
\end{minipage}
\caption{Comparison of the initial (a) and final (b) detector set-up 
during the RHIC 2000 run. Only active elements in central part of the detector 
are shown, more distant elements (trigger counters, TOF and
4 out of 6 ring multiplicity detectors) are not visible.}
\end{figure}

During the first few weeks of RHIC running only a part of the PHOBOS detector
was installed and the magnet was switched off. 
Trigger system was operational from the very beginning
but only a small fraction of silicon sensors 
was present (Fig.\ 1a).
As the highest luminosity was achieved at the end of the RHIC run, almost all
of 3.5 million events collected by PHOBOS were measured with the later detector 
set-up. 

\section{Selection of central events}

The PHOBOS trigger is sensitive both 
to neutral fragments of interacting projectile nuclei registered in ZDC
and to produced charged particles detected by paddle trigger counters.
To reject the background, mostly from the beam-gas interactions, 
we require coincidence of the signals from at least one particle in
each of the two paddle counters.
For the events with low or moderate number of hits in paddle counters
a coincidence of the signals in the ZDC
counters is also demanded. 

Each of 16 scintillator counters in PP and PN registers also
the energy deposited by charged particles and the total signal in paddle
trigger counters is proportional to the number of particles produced
in the Au+Au collision. 
We extract the most central events 
by selecting 6\% of the events with largest 
total signal in PP and PN. For the first physics analysis\ \cite{r1}  
we used only the data measured with the
initial detector setup. Two 
samples of central events, 382 events at $\sqrt{s_{nn}}$ = 56 GeV and 
724 events at $\sqrt{s_{nn}}$ = 130 GeV were analyzed. 

The centrality of the nucleus-nucleus collision can be  
characterized by 
the number of nucleons participating in the interactions. 
Using detailed description of the detector and GEANT based simulation 
program we are able to reproduce satisfactorily 
the detector response for the generated Au+Au collisions 
(from HIJING generator\ \cite{hijing}).
The MC events with largest simulated signal in the paddle 
counters selected by 6\% cut are 
indeed the most central ones.
The distribution of the number of participating nucleons, $N_{part}$, 
for MC events 
is restricted to highest values (Fig.\ 2). We can estimate the mean number
of participating nucleons 
with about 5\% systematic uncertainty (Table\ 1).  \\[0.2cm]
\begin{minipage}{5.8cm}
\epsfxsize=5.8cm\epsfbox{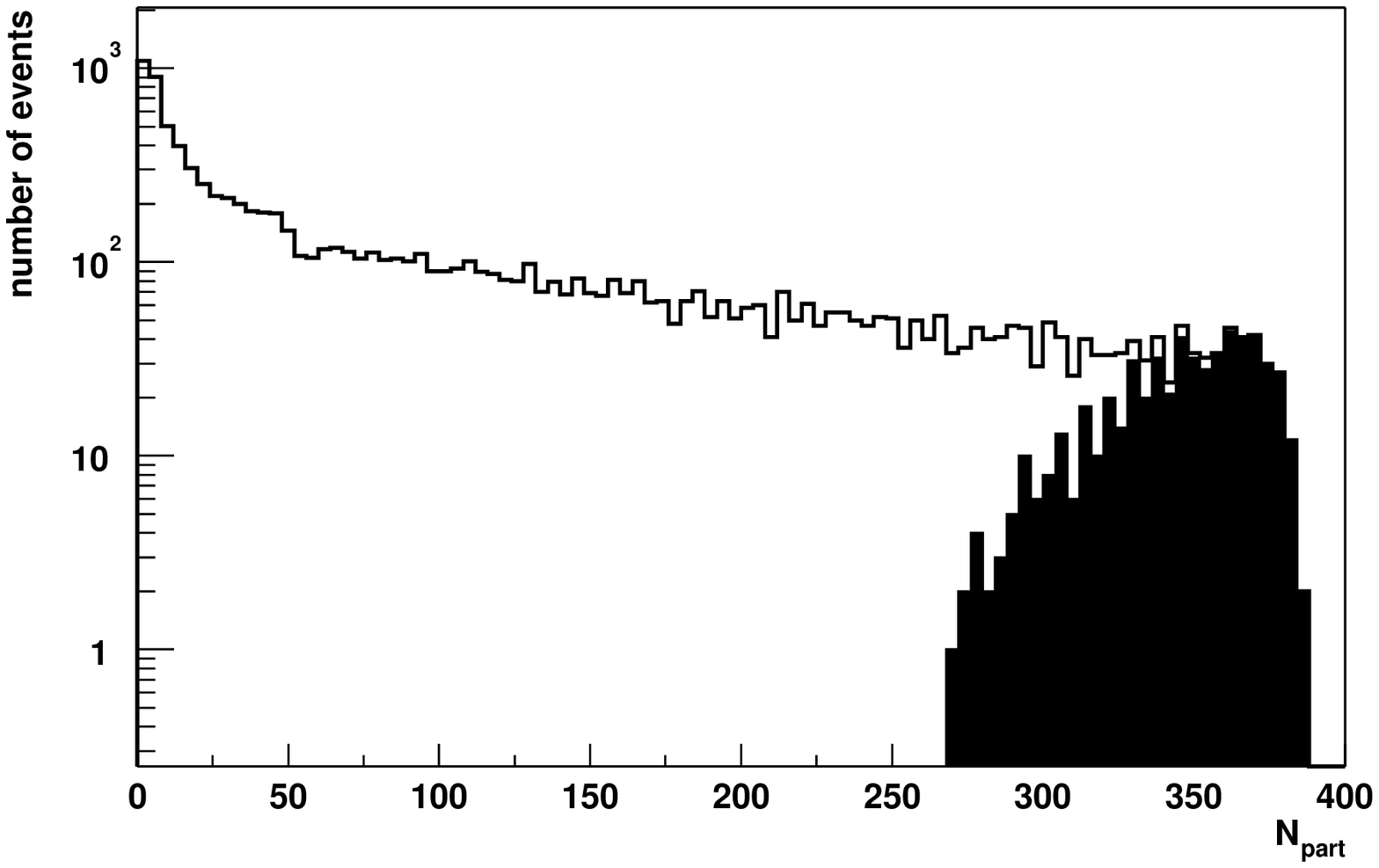}
Figure 2. Distribution of the number of nucleons 
participating in the Au+Au collisions for 
inclusive (histogram) and central HIJING events (shaded area). 
\end{minipage}
\hfill
\begin{minipage}{5.8cm}
\epsfxsize=5.8cm\epsfbox{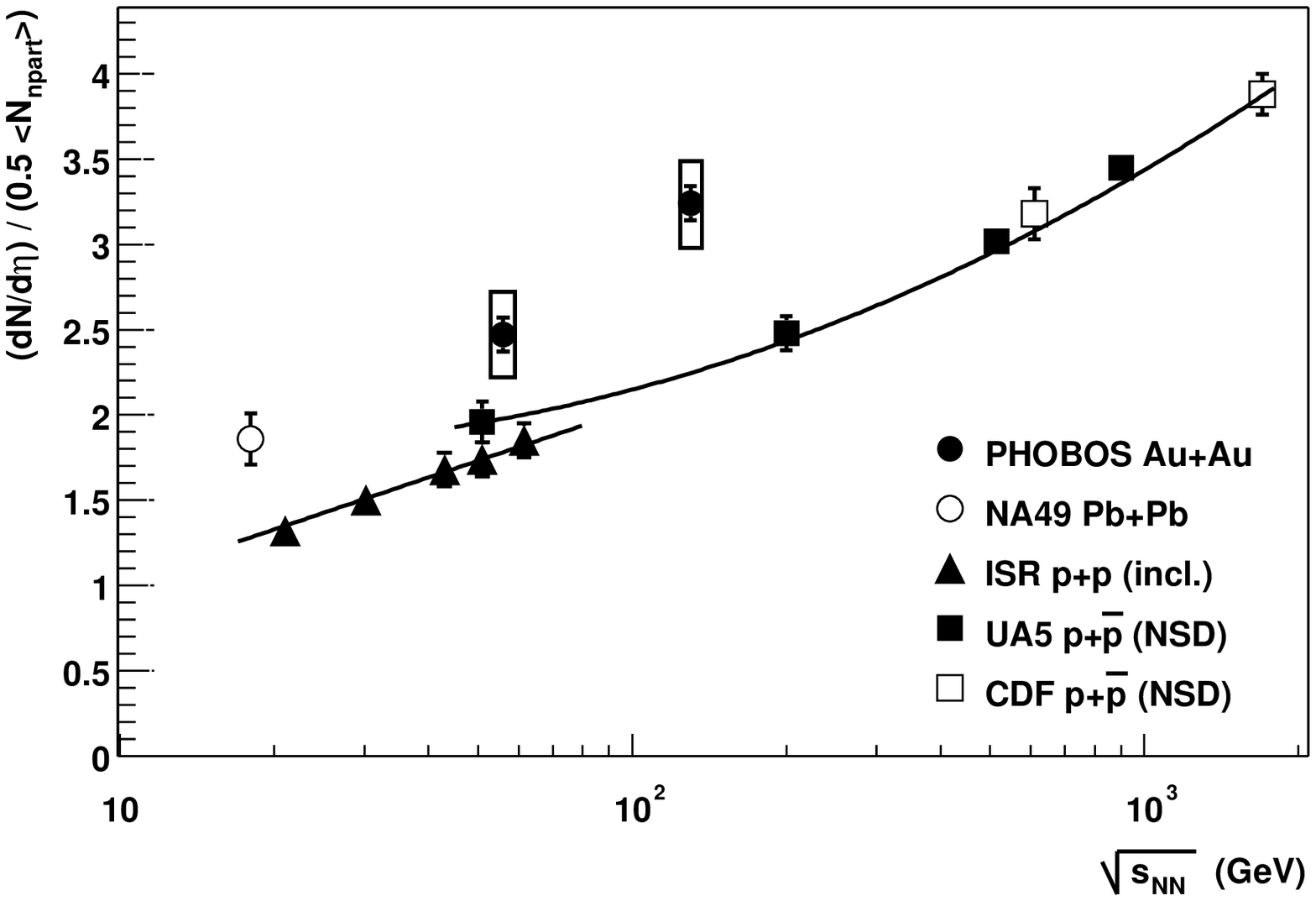}
Figure 3. Charged particles pseudorapidity density normalized per participant 
pair. For PHOBOS systematic uncertainties are shown 
as boxes around data points.
\end{minipage}

\section{Charged particle multiplicity near mid-rapididty}

For both central event samples we performed reconstruction of the straight
line tracks passing at least 4 layers of the spectrometer.
The position of the primary collision vertex
was then determined as the point of closest approach of the found tracks.
We further restricted our analysis to the events with reconstructed
vertex positions within -25$<z_{vtx}<$15 cm from the nominal collision
point along the beam direction.

To measure the charged particle multiplicity,
pairs of hits in two consecutive layers of spectrometer
or vertex detector,
consistent with a track emitted from the collision vertex,  
were formed (tracklets).
%
%
The relation between the number of tracklets and the number 
of primary charged particles is approximately linear and 
was derived applying the same reconstruction
procedure to the MC events processed by GEANT based simulations
of PHOBOS detector. The proportionality factors $\alpha(z_{vtx})$ 
agree better than 5\% for the MC events obtained from HIJING and 
VENUS\ \cite{venus}, eventhough the total multiplicities 
in these models differ by a factor 2-3. 
The extracted values of $\alpha(z_{vtx})$ contain corrections for 
the tracklets finding inefficiency, contributions 
from secondary particles, decay products and noise hits, stopping of 
primary particles in the beam pipe and, the most important, 
geometrical acceptance.
Overall systematic uncertainty of $\alpha(z_{vtx})$
is less than 8\%.
Integrated values of the charged particle pseudorapidity density in 
the $\eta$ range -1\ to\ 1, $dN/d\eta|_{|\eta|<1}$, are given in Table 1.
The densities normalized per pair of participant nucleons, 
$(dN/d\eta|_{|\eta|<1})/(0.5<N_{part}>)$ are directly compared to 
results obtained in other experiments\ \cite{na49,isr,pantip} (Fig.\ 3).
For central Au+Au collisions we observe a significantly larger
particle density per participant pair than for 
$pp$ or $p\overline{p}$ interactions 
at similar energies. The increase of the
particle density in nucleus-nucleus collisions 
with energy is also faster than in 
elementary interactions. 
At RHIC energies, 
$\sqrt{s_{nn}}$ = 56 GeV and 130 GeV, this increase is 31\%.
The presented  results impose restrictions on  
particle production models, ruling out simple superposition 
models\ \cite{wounded}.

\begin{table}[t]
\caption{Charged particle pseudorapidity density for central events.}
\vspace{0.2cm}
\begin{center}
\begin{tabular}{|c|c|c|}
\hline
  \raisebox{0pt}[10pt][5pt]{}  &  $\sqrt{s_{nn}}$ = 56 GeV  &  $\sqrt{s_{nn}}$ = 130 GeV \\
\hline
 \raisebox{0pt}[12pt][6pt]{} $<N_{part}>$ 
      &  330 $\pm$4 (stat.) $^{+10}_{-15}$ (syst.)
          &   343 $\pm$4 (stat.) $^{+7}_{-14}$ (syst.) \\
\hline
 \raisebox{0pt}[12pt][6pt]{} $dN/d\eta|_{|\eta|<1}$ 
      &  408 $\pm$12 (stat.) $\pm$30 (syst.)
          &   555 $\pm$12 (stat.) $\pm$35 (syst.)  \\
\hline
 \raisebox{0pt}[12pt][6pt]{} $\frac{dN/d\eta|_{|\eta|<1}}{0.5<N_{part}>}$
      &  2.47 $\pm$0.1(stat.) $\pm$0.25(syst.)
          &   3.24 $\pm$0.1(stat.) $\pm$0.25(syst.)    \\
\hline
\end{tabular}
\end{center}
\end{table}

\section{Work in progress}

The first results presented in the previous section were obtained 
using only a small fraction of the data collected by PHOBOS. Analysis of the 
full data set collected later with the complete detector set-up 
will provide much more detailed information 
on the properties of Au+Au collisions. The normalized charged particle
densities can be measured in a wide range of $N_{part}$, 
also for semi-peripheral collisions for which some models give 
significantly different
predictions\ \cite{satur} (Fig.\ 4). With a complete arm N of spectrometer 
placed in the magnetic field we can measure 
particle momenta and identify them (Fig.\ 5) and thus obtain 
anti-particles to particles ratios.
Using full multiplicity detector 
we can measure pseudorapidity density of primary particles in 
11 rapidity units. Data from multiplicity detector will be used to 
study elliptic flow of charged particles and fluctuations 
in angular distribution. \\
\begin{minipage}{5.8cm}
\epsfxsize=5.8cm\epsfbox{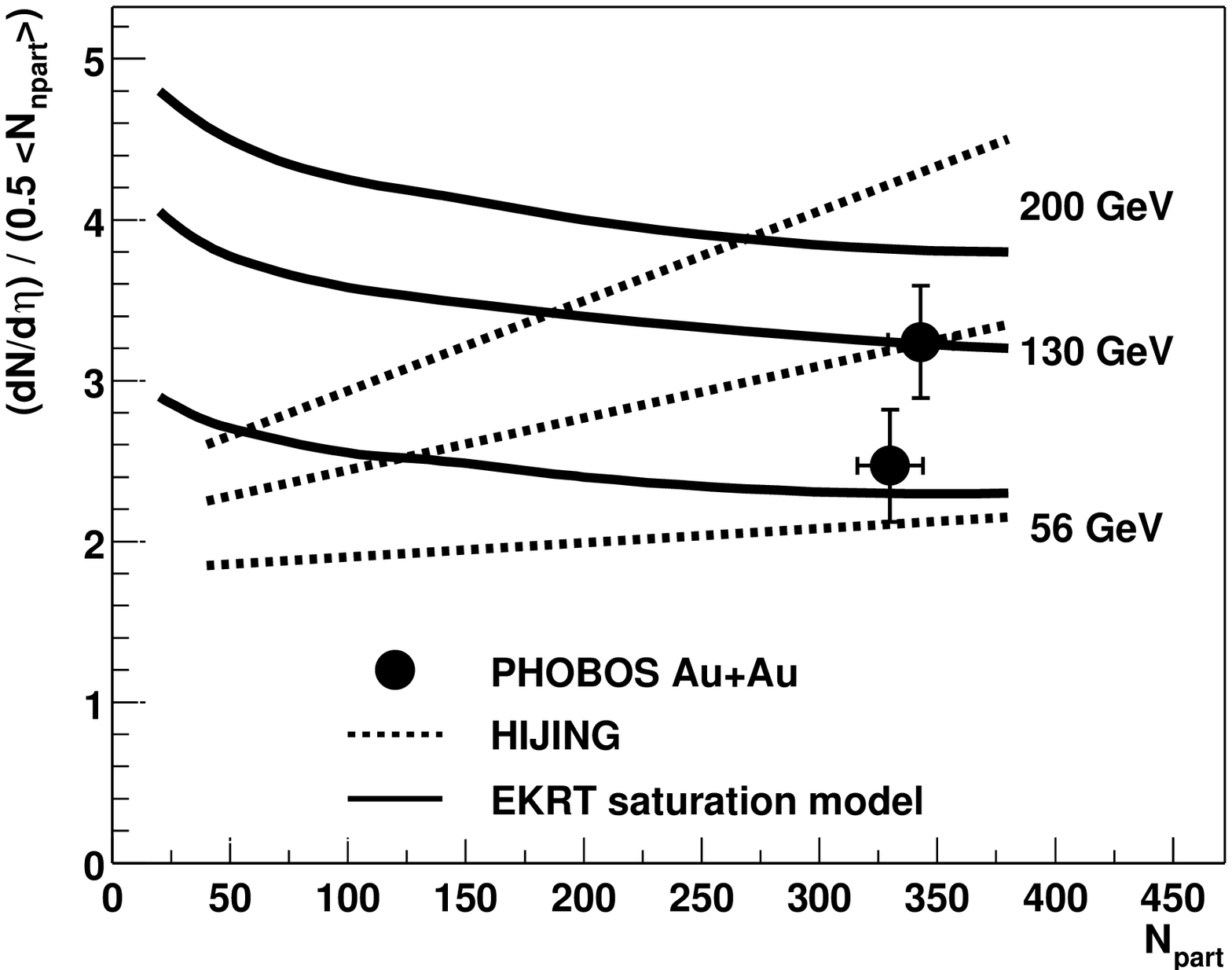}
Figure 4. Dependence of the charged particle pseudorapidity density on the
centrality of the Au+Au collision for HIJING model and EKRT saturation model.
\end{minipage}
\hfill
\begin{minipage}{5.8cm}
\vspace*{0.25cm}
\epsfxsize=5.8cm\epsfbox{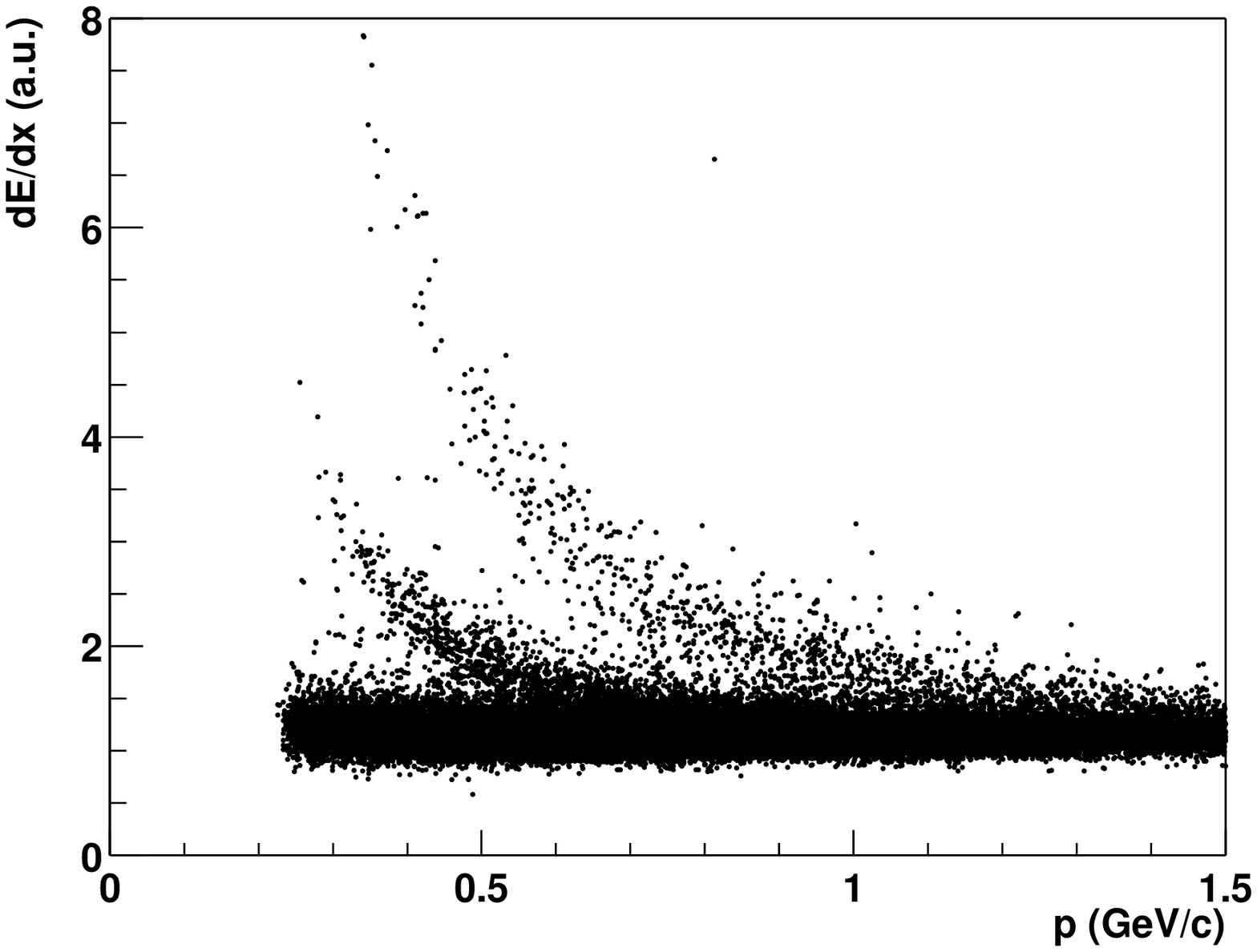}
Figure 5. Identification of the particles in PHOBOS spectrometer.
Proton and kaon bands 
start to separate from horizontal pions band
 at about 1.1 and 0.6 GeV/c respectively.
\end{minipage}

\section*{Acknowledgments}

We acknowledge the generous support of the entire RHIC project personnel, C-A
and Chemistry Departments at BNL. We thank Fermilab and CERN for help in
silicon detector assembly. We thank the MIT School of Science and LNS for
financial
support. This work was partially supported by US DoE grants 
\mbox{DE-AC02-}\-\mbox{98CH10886},
\mbox{DE-FG02-}\-\mbox{93ER40802}, 
DE-FC02-94ER40818, 
\mbox{DE-FG02}\-\mbox{94ER40865}, 
DE-FG02-99ER41099, W-31-109-ENG-38.
NSF grants 9603486, 9722606 and 0072204. The Polish group 
from INP was partially supported by KBN grant 2 P03B
04916. The NCU group was partially supported by NSC of Taiwan under
contract NSC 89-2112-M-008-024.


\section*{References}
\begin{thebibliography}{99}
\bibitem{phobos1} B.Back {\it et al.}, (PHOBOS), Nucl. Phys. {\bf A661} (1999) 690.
\bibitem{r1} B.B.Back {\it at al.}, (PHOBOS), 
Phys. Rev. Lett. {\bf 85} (2000) 3100. 
\bibitem{hijing} M.Gyulassy and X.N.Wang, Phys. Rev. {\bf D44} (1991) 3501.
\bibitem{venus} K.Werner, Phys.\ Rep.\ {\bf 232} (1993) 87.
\bibitem{na49} J.B\"achler et al., Nucl.\ Phys.\ {\bf A661} (1999) 45.
\bibitem{isr} W.Thome et al., Nucl. Phys., B129 (1977) 365. 
\bibitem{pantip} F.Abe et al., Phys. Rev.\ {\bf D41} (1990) 2330.
\bibitem{wounded} A.Bia\l as, B.Bleszy\'{n}ski and W.Czy\.{z}, Nucl.\ Phys.\ {\bf B111} (1976) 461.
\bibitem{satur} X.N.Wang and M.Guylassy, nucl-th/0008014 (2000).

\end{thebibliography}
\end{document}